\documentclass[twoside]{article}
\usepackage{Proc_NTSE_18}
\usepackage{bebands-ntse18}

\begin{document}
\thispagestyle{plain}

% proceedings publication info footer
\setcounter{page}{250}
\publref{Caprio}  % http://www.ntse.khb.ru/files/uploads/2018/proceedings/Caprio.pdf
\setcounter{page}{1}

% spacer between footer and author footnote
\begingroup
\makeatletter
\renewcommand{\@makefnmark}{\raisebox{0pt}[0pt][0pt]{}}
\footnotetext{}
\makeatother
\endgroup

% set up for ``Present address'' footnote
% set footnote symbol type so distinct from text footnotes
\newcommand{\butlerpresentaddress}{\textit{Present address:} Department of
  Physics and Astronomy, Michigan State University, East Lansing, MI 48824,
  USA.}
\renewcommand{\thefootnote}{\fnsymbol{footnote}}

\begin{center}
{\Large \bf \strut
Robust \textit{ab initio} predictions for nuclear rotational structure in the \boldmath$\isotope{Be}$ isotopes
\strut}\\
\vspace{10mm}
{\large \bf 
M.~A.~Caprio$^{a}$, P.~J.~Fasano$^{a}$, J.~P.~Vary$^{b}$, P.~Maris{$^b$} and J.~Hartley$^{a,c,}$\footnote{\butlerpresentaddress}}
\end{center}

\noindent
{\small $^a$\it Department of Physics, University of Notre Dame, Notre Dame, Indiana 46556, USA} \\
{\small $^b$\it Department of Physics and Astronomy, Iowa State University, Ames, Iowa 50011, USA} \\
{\small $^c$\it Department of Chemistry and Physics, Erskine College, Due West, South Carolina 29639, USA}  % 2 Washington St

\markboth
{M.~A.~Caprio, P.~J.~Fasano, J.~P.~Vary, P.~Maris and J.~Butler}
{Robust \textit{ab initio} predictions for nuclear rotational structure} 

\begin{abstract}
No-core configuration interaction (NCCI) calculations for $p$-shell nuclei give
rise to rotational bands, identified by strong intraband $E2$ transitions and by
rotational patterns for excitation energies, electromagnetic moments, and
electromagnetic transitions.  However, convergence rates differ significantly
for different rotational observables and for different rotational bands.  The
choice of internucleon interaction may also substantially impact the convergence
rates.  Consequently, there is a substantial gap between simply observing the
\textit{qualitative} emergence of rotation in \textit{ab initio} calculations
and actually carrying out detailed \textit{quantitative} comparisons.  In this
contribution, we illustrate the convergence properties of rotational band energy
parameters extracted from NCCI calculations, and compare these predictions with experiment, for the isotopes
$\isotope[7\text{--}11]{Be}$, and for the JISP16 and Daejeon16 interactions.
\\[\baselineskip]
\end{abstract}

\setcounter{footnote}{0}
\renewcommand{\thefootnote}{\arabic{footnote}}  % reset for footnotes to default format for main text

%%%%%%%%%%%%%%%%%%%%%%%%%%%%%%%%%%%%%%%%%%%%%%%%%%%%%%%%%%%%%%%%

\section{Introduction}

\textit{Ab initio} nuclear theory aims to describe nuclei, with quantitative
precision, from the underlying internucleon interactions.  Light nuclei are
known to display rotational band structure (\textit{e.g.},
Refs.~\cite{rogers1965:nonspherical-nuclei,vonoertzen1996:be-molecular,vonoertzen1997:be-alpha-rotational,freer2007:cluster-structures}).
Therefore, we should at least aspire for \textit{ab initio} theory to be able to
predict rotational band structure.  However, there are challenges to obtaining
converged calculations of the relevant observables, both energies and
electromagnetic transition
strengths~\cite{pervin2007:qmc-matrix-elements-a6-7,bogner2008:ncsm-converg-2N,cockrell2012:li-ncfc,maris2013:ncsm-pshell}.

There are thus a few basic questions to be asked about the emergence of
rotation in \textit{ab initio} calculations of light nuclei:

(1)~Is there a \textit{qualitative emergence} of rotational ``features'' in the
calculated results?  These features include rotational energy patterns and transition
patterns.

(2)~Can robust \textit{quantitative predictions} be made for rotational
observables?  These observables include rotational band energy parameters or
intrinsic matrix elements.  Here we must have good convergence of the results
of the many-body calculation, at which point we can then explore the robustness
of the predictions across possible internucleon interactions.

(3)~Once the \textit{ab initio} description for nuclear rotation is solidly
established, what can it tell us about the structure of these rotational states?
This understanding may come in the form of identifying, \textit{e.g.}, many-body
symmetries~\cite{dytrych2007:sp-ncsm-dominance,dytrych2013:su3ncsm,johnson2015:spin-orbit,mccoy2018:spncci-busteni17}
or cluster
structure~\cite{vonoertzen1996:be-molecular,freer2007:cluster-structures,kanadaenyo2012:amd-cluster}
underlying the rotation.

Regarding the first, qualitative question, no-core configuration interaction
(NCCI)~\cite{barrett2013:ncsm} calculations for $p$-shell nuclei give rise to
rotational bands, identified by strong intraband $E2$ transitions and by
rotational patterns for excitation energies, electromagnetic moments, and
electromagnetic
transitions~\cite{caprio2013:berotor,maris2015:berotor2-WITH-ERRATUM}
(see also Ref.~\cite{caprio2015:berotor-ijmpe} for a pedagogical review).
However, convergence rates differ significantly for different rotational
observables and for different rotational bands, as well as in calculations based
on different internucleon interactions~\cite{caprio2015:berotor-ijmpe}.
Consequently, there is a substantial gap between simply observing the
\textit{qualitative} emergence of rotation in \textit{ab initio} calculations
and actually obtaining detailed \textit{quantitative} predictions for comparison
with experiment.

In this contribution, we focus on quantitative predictions of rotational band
energy parameters.  We first illustrate the convergence properties of rotational
parameters extracted from NCCI calculations, taking $\isotope[11]{Be}$ as an
example (Sec.~\ref{sec:illustration}).  We then obtain \textit{ab initio}
predictions for rotational band parameters across the isotopes
$\isotope[7\text{--}11]{Be}$.  We explore the robustness of these predictions
with respect to the choice of internucleon interaction
(JISP16~\cite{shirokov2007:nn-jisp16} and
Daejeon16~\cite{shirokov2016:nn-daejeon16}) and compare these predictions with
experiment (Sec.~\ref{sec:params}).

%%%%%%%%%%%%%%%%%%%%%%%%%%%%%%%%%%%%%%%%%%%%%%%%%%%%%%%%%%%%%%%%

\section{Illustration: Rotational bands in \boldmath$\isotope[11]{Be}$}
\label{sec:illustration}

\subsection{Excitation spectrum and bands}
\label{sec:illustration:spectrum}
%%%%%%%%%%%%%%%%%%%%%%%%%%%%%%%%%%%%%%%%%%%%%%%%%%%%%%%%%%%%%%%%
\begin{figure}[t]
\centerline{\includegraphics[width=0.75\textwidth]{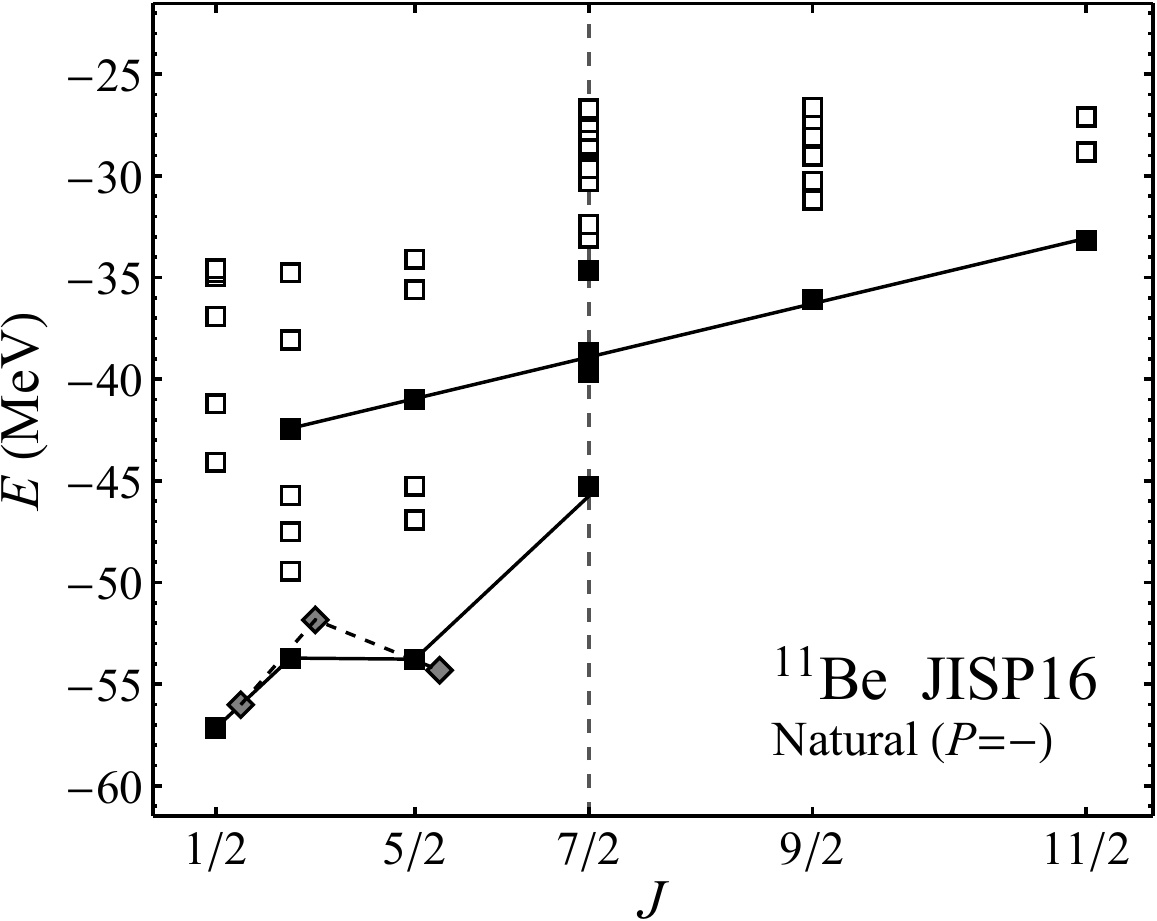}}
\caption{Calculated energy eigenvalues (squares) for states in the natural
  (negative) parity space of $\isotope[11]{Be}$, with the JISP16 interaction; the three lowest calculated unnatural
  (positive) parity states are also shown (diamonds, displaced
  horizontally for clarity).  
  Energies are plotted with respect to angular momenta scaled as $J(J+1)$.  Solid symbols indicate band
  members, as identified by strong $E2$ transitions and other supporting
  observables.  Lines indicate rotational fits~(\ref{eqn-EJ-stagger}) to the calculated
  energies of the band members.  Calculated with $\Nmax=8$ (or $\Nmax=9$
  for unnatural parity) at $\hw=20\,\MeV$.}
\label{fig:11be1-levels}
\end{figure}
%%%%%%%%%%%%%%%%%%%%%%%%%%%%%%%%%%%%%%%%%%%%%%%%%%%%%%%%%%%%%%%%

To illustrate the nature of the rotational bands obtained in NCCI calculations,
let us take $\isotope[11]{Be}$ as an example.  In this nucleus, we encounter
bands with qualitatively different termination and convergence properties.

A calculated eigenvalue spectrum for $\isotope[11]{Be}$ is shown in
Fig.~\ref{fig:11be1-levels}.\footnote{The NCCI calculations shown here are
  obtained using the code
  MFDn~\cite{maris2010:ncsm-mfdn-iccs10,aktulga2013:mfdn-scalability,shao2018:ncci-preconditioned}.}
The detailed results depend upon the particular choice of the internucleon
interaction (here, JISP16~\cite{shirokov2007:nn-jisp16} plus Coulomb interaction
between protons) and truncated space (here, up to $\Nmax=8$ excitation quanta,
and with oscillator basis length scale given by $\hw=20\,\MeV$), as we shall
explore in subsequent sections, but the example calculation in
Fig.~\ref{fig:11be1-levels} provide a representative illustration of the general
rotational features.

Band members are expected to have energies following the rotational formula
$E(J)=E_0+AJ(J+1)$, where the rotational energy constant
$A\equiv\hbar^2/(2\cal{J})$ is inversely related to the moment of inertia
$\cal{J}$ of the rotational intrinsic state, and the intercept parameter
$E_0=E_K-AK^2$ is related to the energy $E_K$ of the rotational intrinsic
state~\cite{ring1980-nuclear-many-body,rowe2010:collective-motion}.\footnote{Under the
assumption of axial symmetry, each band is characterized by a projection $K$ of
the angular momentum on the intrinsic symmetry axis, and the rotational band members have angular momenta $J\geq K$.}  The level energies in
Fig.~\ref{fig:11be1-levels} are therefore plotted against angular momenta scaled
as $J(J+1)$, so that the energies within a band follow a linear pattern.  For
$K=1/2$ bands, the Coriolis contribution to the kinetic energy significantly
modifies this pattern, yielding an energy staggering which is given, in
first-order perturbation theory, by
\begin{equation}
\label{eqn-EJ-stagger}
E(J)=E_0+A\bigl[J(J+1)+
a(-)^{J+1/2}(J+\tfrac12)
%%\underbrace{a(-)^{J+1/2}(J+\tfrac12)}_{\text{Coriolis ($K=1/2$)}}
\bigr],
\end{equation}
where the Coriolis decoupling parameter $a$ depends upon the structure
of the rotational intrinsic state.

Rotational band members are shown in Fig.~\ref{fig:11be1-levels} by filled
symbols.  These identifications are based not simply on the level energies, but
rather on strong $E2$ connections (for illustration, see Figs.~6, 10, and 14 of
Ref.~\cite{caprio2015:berotor-ijmpe}).

The lowest filling of harmonic oscillator shells possible for
$\isotope[11]{Be}$, consistent with Pauli exclusion, has an odd number of
nucleons in the negative-parity $p$ shell.  Thus, the ``natural'' parity for
$\isotope[11]{Be}$, as would be obtained in a traditional $0\hw$ shell model
description or an $\Nmax=0$ NCCI calculation, is negative.  In
Fig.~\ref{fig:11be1-levels}, we focus on the natural (negative) parity states
(indicated by squares) and show only the lowest three ``unnatural'' (positive)
parity states for comparison (diamonds).

In this particular calculation (Fig.~\ref{fig:11be1-levels}), the lowest
positive parity state ($1/2^+$) lies slightly above the lowest negative parity
state ($1/2^-$).  However, experimentally, the ground state of
$\isotope[11]{Be}$ is $1/2^+$, lying $0.320\,\MeV$ below a $1/2^-$ excited
state~\cite{npa2012:011}.  (Such a reversal of the ground state parity relative
to the natural parity is known as \textit{parity inversion}.)  Different rates
of convergence between the natural and unnatural parity states makes it
challenging to predict the level ordering when the separation of energies is so
small.

The lowest negative-parity band has $K^P=1/2^-$ and apparently terminates with the
$7/2^-$ state.  This angular momentum $J=7/2$ (indicated by the dashed vertical
line in Fig.~\ref{fig:11be1-levels}) is the highest which can be obtained in a
$p$-shell description of $\isotope[11]{Be}$, that is, in the shell model $0\hw$
valence space or in an NCCI $\Nmax=0$ calculation.

On the other hand, the excited negative-parity $K^P=3/2^-$ band extends past the
maximal valence angular momentum.  The $J\leq7/2$ band members lie in a region
of the excitation spectrum with a comparatively high level density and are thus
subject to mixing with the ``background'' non-rotational states.  Such mixing
occurs when an approximate accidental degeneracy of the rotational state and
background states leads to a small energy denominator for mixing.  Since we found that the
energies of these states converge differently with $\Nmax$ and $\hw$, mixing for
any given rotational state might arise in one truncated calculation but not the
next.  For instance, in the particular calculation shown here, the $E2$
strengths suggest that the excited $7/2^-$ band member is actually fragmented
over three states, as indicated by the filled symbols.  Starting with $J=9/2$,
this band becomes yrast, and the band members are comparatively well-isolated.

The lowest calculated positive parity states are the $1/2^+$, $3/2^+$, and
$5/2^+$ members of a $K^P=1/2^+$ band.  This band continues to much higher
angular momentum than shown here, as may be seen in Fig.~3(e) of
Ref.~\cite{maris2015:berotor2-WITH-ERRATUM}.

\subsection{Dependence of the calculated bands on \boldmath$\Nmax$ truncation}
\label{sec:illustration:nmax}
%%%%%%%%%%%%%%%%%%%%%%%%%%%%%%%%%%%%%%%%%%%%%%%%%%%%%%%%%%%%%%%%
\begin{figure}[tb]
\centerline{\includegraphics[width=0.75\textwidth]{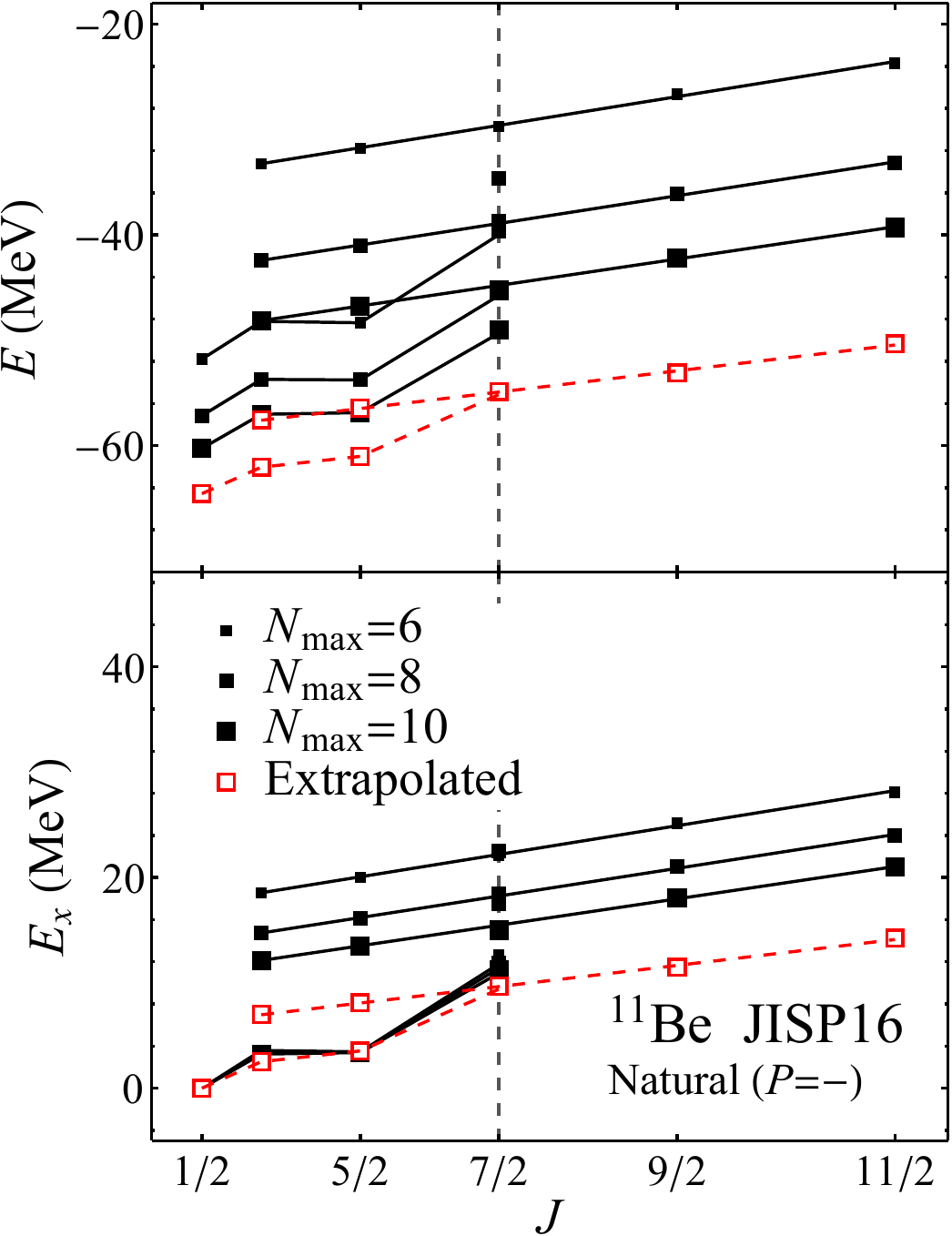}}
\caption{Convergence of calculated energy eigenvalues~(top) and excitation
  energies~(bottom) with $\Nmax$, for rotational band members in the natural
  (negative) parity space of $\isotope[11]{Be}$.  Successively larger symbols
  indicate successively higher $\Nmax$ values ($\Nmax=6$, $8$, and $10$).  The
  open symbols indicate exponentially extrapolated level energies.  Lines
  indicate rotational fits~(\ref{eqn-EJ-stagger}) to the calculated (or
  extrapolated) energies of these band members.  Calculated with the JISP16
  interaction at $\hw=20\,\MeV$.}
\label{fig:11be1-energies-nmax}
\end{figure}
%%%%%%%%%%%%%%%%%%%%%%%%%%%%%%%%%%%%%%%%%%%%%%%%%%%%%%%%%%%%%%%%

While Fig.~\ref{fig:11be1-levels} illustrates the qualitative features of the
rotational patterns which arise in NCCI calculations, it represents an
approximate calculation of the spectrum, as obtained in a truncated space.  It
is thus only an unconverged ``snapshot'', along the path towards the true
results which would be obtained if the many-body problem could be solved in the
full, untruncated many-body space.

To see how the rotational pattern evolves, as we progress through calculations
truncated to successively higher numbers of oscillator excitations, let us focus
on the rotational band members in the negative parity space of
$\isotope[11]{Be}$.  We trace out the energies obtained for $\Nmax=6$, $8$, and
$10$ in Fig.~\ref{fig:11be1-energies-nmax}~(top).  These energies are far from
converged.  Each level moves downward by several $\MeV$ for each step in
$\Nmax$.

However, the energies of levels within a band move downward
nearly in lockstep.
Thus, if we look instead at \textit{excitation} energies, as in
Fig.~\ref{fig:11be1-energies-nmax}~(bottom), here taken relative to the lowest
($1/2^-$) negative parity state, the energies of the $K^P=1/2^-$ band members
are comparatively stable.  In fact, only the excitation energy of the
terminating $7/2^-$ band member changes noticeably at an $\MeV$ scale.

The $K^P=3/2^-$ band is still converging downward relative to the
$K^P=1/2^-$ band with increasing $\Nmax$, reflected in the decreasing excitation
energies in Fig.~\ref{fig:11be1-energies-nmax}~(bottom).  It is not obvious
where we could expect these excitation energies to settle, if we could solve
the nuclear many-body problem in the full, untruncated space.

However, we can
attempt to \textit{estimate} the full-space result by assuming a functional form
for the convergence of the calculated energy eigenvalues.  For instance, the
sequence of eigenvalues computed at successive $\Nmax$ appears to follow a
roughly geometric convergence pattern, suggestive of a decaying exponential in
$\Nmax$~\cite{bogner2008:ncsm-converg-2N,forssen2008:ncsm-sequences,maris2009:ncfc}:
\begin{equation}
\label{eqn-exp}
E(\Nmax)=c_0+c_1 \exp(-c_2\Nmax).
\end{equation}
Since calculated energies at three $\Nmax$ values are required to fix the three
parameters in~(\ref{eqn-exp}), this functional form provides a three-point
extrapolation formula for energies, giving the estimate $E\rightarrow c_0$ as
$\Nmax\rightarrow\infty$.  This is only an \textit{ad hoc} phenomenological
prescription, but it provides an idea of what might be plausible for the
full-space results.

Extrapolated energies for the $\isotope[11]{Be}$ band members are shown in
Fig.~\ref{fig:11be1-energies-nmax} (open symbols): as eigenvalues~(top), and
then as excitation energies, taken relative to the extrapolated $1/2^-$
eigenvalue (bottom).  While the extrapolated energies of the $K^P=3/2^-$ band
members still lie above those of the $K^P=1/2^-$ band at lower angular momenta,
the lower slope of the excited band, combined with the Coriolis staggering of
the $K^P=1/2^-$ band members, leads to nearly degenerate extrapolated energies
for the $7/2^-$ members of these two bands.  If such a degeneracy were to arise,
we could expect significant two-state mixing to occur between the two rotational
configurations in the $7/2^-$ band members (similar to the mixing of the excited
$7/2^-$ with the background states seen already at higher excitation energy, in
Fig.~\ref{fig:11be1-levels}).  The level repulsion induced by this mixing would
be highly non-perturbative and would thus frustrate any simple attempt at
extrapolating the energies from low-$\Nmax$ calculations where the mixing is not
yet in effect.

\subsection{Stability of calculated rotational energy parameters}
\label{sec:illustration:params}
%%%%%%%%%%%%%%%%%%%%%%%%%%%%%%%%%%%%%%%%%%%%%%%%%%%%%%%%%%%%%%%%
\begin{figure}[t]
\centerline{\includegraphics[width=1.00\textwidth]{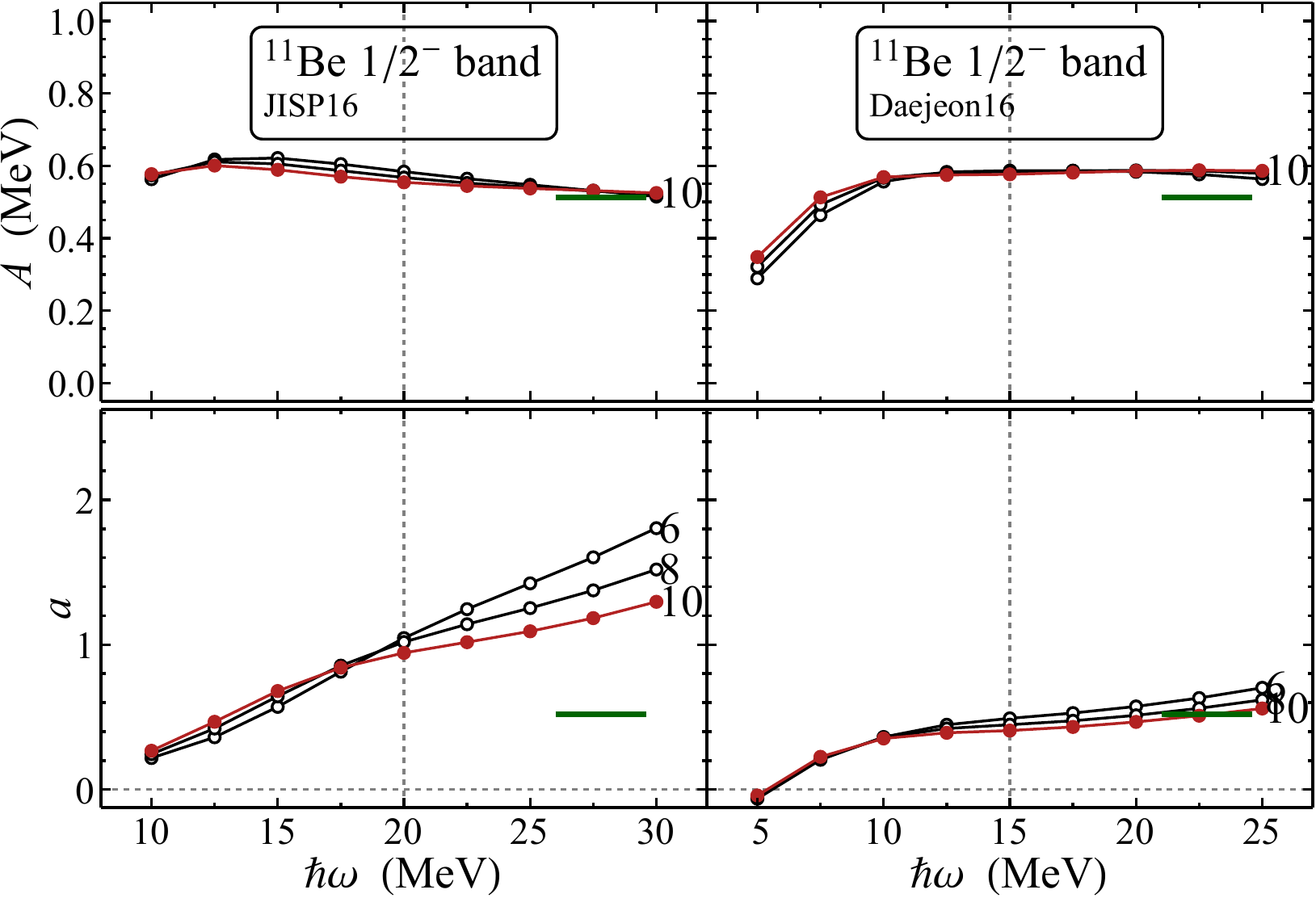}}
\caption{Dependence of the extracted rotational energy parameters, for the
  $K^P=1/2^-$ band of $\isotope[11]{Be}$, on the truncation parameters $\Nmax$
  and $\hw$ of the NCCI space in which the calculations are carried out.
  Successive curves are for successively higher $\Nmax$ values ($\Nmax=6$,
  $8$, and $10$, noted alongside curve).  Experimental values
  (horizontal lines) are shown for comparison ($A=0.51\,\MeV$ and $a=0.52$).
  The vertical dashed lines indicate the approximate location of the variational
  energy minimum, in $\hw$, of the calculated ground state energy (see text).}
\label{fig:11be1-band1-params-scan}
\end{figure}
%%%%%%%%%%%%%%%%%%%%%%%%%%%%%%%%%%%%%%%%%%%%%%%%%%%%%%%%%%%%%%%%

%%%%%%%%%%%%%%%%%%%%%%%%%%%%%%%%%%%%%%%%%%%%%%%%%%%%%%%%%%%%%%%%
\begin{figure}[tb]
\centerline{\includegraphics[width=1.00\textwidth]{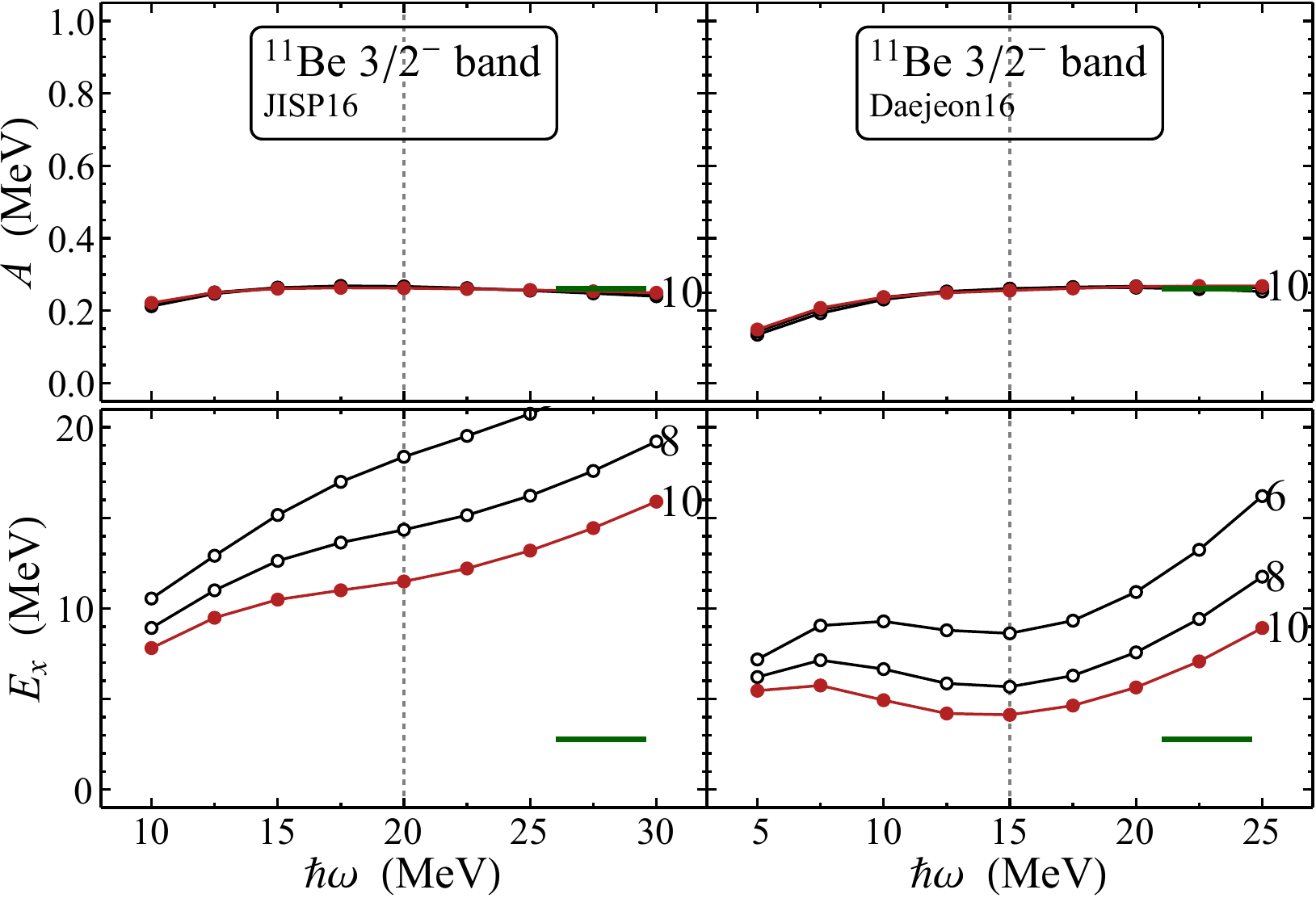}}
\caption{Dependence of the extracted rotational energy parameters, for the
  excited $K^P=3/2^-$ band of $\isotope[11]{Be}$, on the truncation parameters $\Nmax$ and
  $\hw$ of the NCCI space in which the calculations are carried out.
  Successive curves are for successively higher $\Nmax$ values ($\Nmax=6$,
  $8$, and $10$, noted alongside curve).  
  The band excitation energy $E_x$ is taken relative to the $K^P=1/2^-$ band.
Experimental values (horizontal lines) are shown for comparison
  ($A=0.26\,\MeV$ and $E_x=2.77\,\MeV$).
  The vertical dashed lines indicate the approximate location of the variational
  energy minimum, in $\hw$, of the calculated ground state energy (see text).
}
\label{fig:11be1-band2-params-scan}
\end{figure}
%%%%%%%%%%%%%%%%%%%%%%%%%%%%%%%%%%%%%%%%%%%%%%%%%%%%%%%%%%%%%%%%

%%%%%%%%%%%%%%%%%%%%%%%%%%%%%%%%%%%%%%%%%%%%%%%%%%%%%%%%%%%%%%%%
\begin{figure}[p]
\centerline{\includegraphics[width=1.00\textwidth]{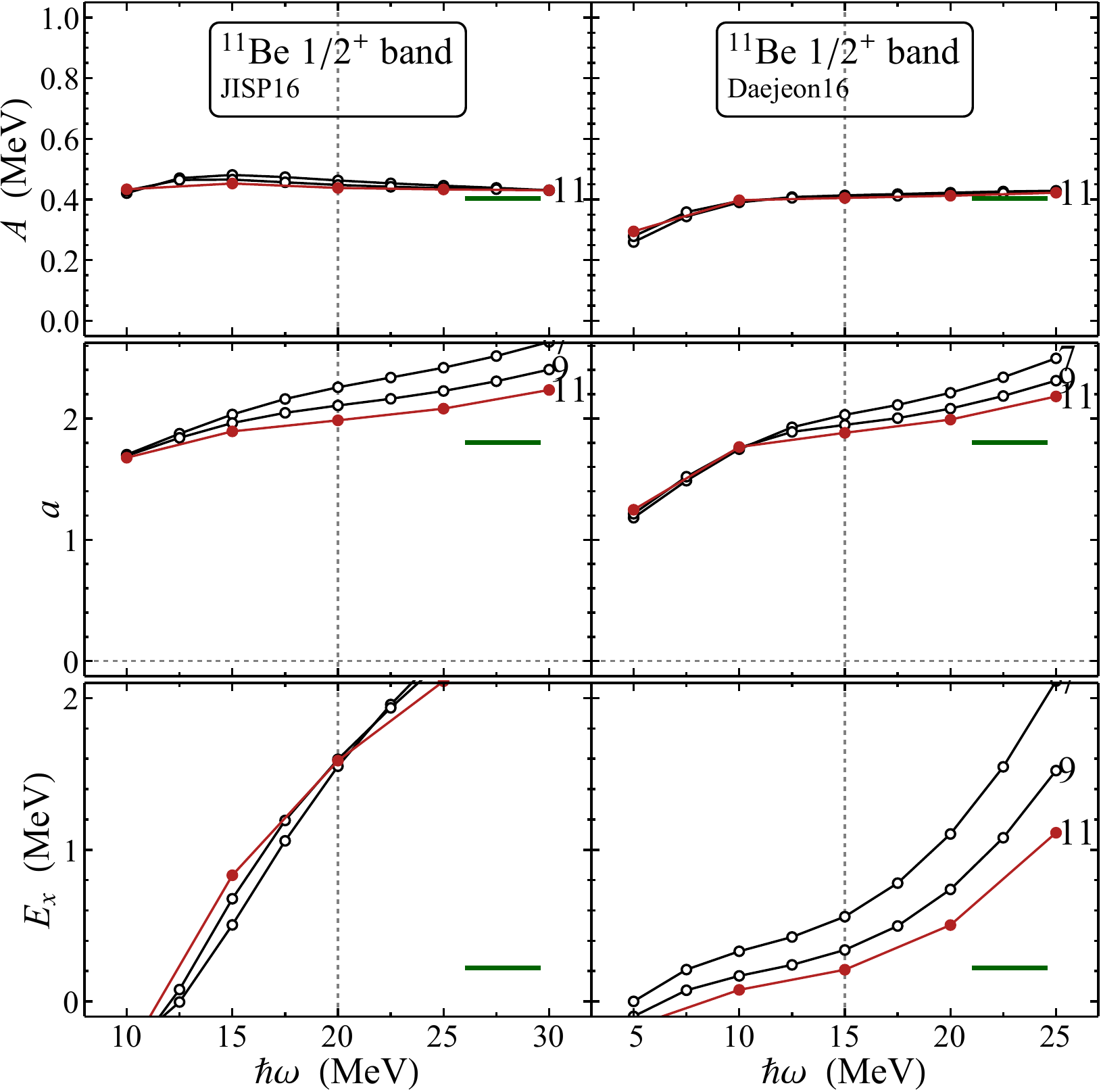}}
\caption{Dependence of the extracted rotational energy parameters, for the
  unnatural-parity $K^P=1/2^+$ band of $\isotope[11]{Be}$, on the truncation parameters $\Nmax$ and
  $\hw$ of the NCCI space in which the calculations are carried out.
  Successive curves are for successively higher $\Nmax$ values ($\Nmax=7$,
  $9$, and $11$, noted alongside curve).
  The band excitation energy $E_x$ is taken relative to the $K^P=1/2^-$ band.
  Experimental values (horizontal lines) are shown for comparison
  ($A=0.40\,\MeV$, $a=1.80$, and $E_x=0.22\,\MeV$).
  The vertical dashed lines indicate the approximate location of the variational
  energy minimum, in $\hw$, of the calculated ground state energy (see text).
}
\label{fig:11be0-band1-params-scan}
\end{figure}
%%%%%%%%%%%%%%%%%%%%%%%%%%%%%%%%%%%%%%%%%%%%%%%%%%%%%%%%%%%%%%%%

Rotational energy parameters extracted from calculations for the
$\isotope[11]{Be}$ bands are examined, as functions of $\Nmax$ and $\hw$ and for
different interactions, in
Figs.~\ref{fig:11be1-band1-params-scan}--\ref{fig:11be0-band1-params-scan}.
There are several questions to be answered for these extracted parameter values:

(1)~Are the calculated values stable against the parameters $\Nmax$ and $\hw$ of
the truncated space?

(2)~If so, are the predictions consistent across the
different internucleon interactions?

(3)~How do these predictions then compare
to experiment?

Recall that these parameters are the inertial (or slope) parameter $A$, energy
(or intercept) parameter $E_0$, and Coriolis decoupling (or staggering)
parameter $a$ (for $K=1/2$).  The excitation energy $E_x$ of bands relative to
each other is then measured by the difference in their band energy parameters
$E_0$ (we use the $K^P=1/2^-$ band as our reference for excitation
energies).\footnote{Translating differences of band energy parameters into
  differences in intrinsic excitation energies would require that we also take into
  account the correction $\propto K^2$ (Sec.~\ref{sec:illustration:spectrum}).}

It is instructive to examine and compare the convergence behaviors of the
parameters $A$, $a$, and $E_x$ for the various bands, and subject to different
interactions.  Successive curves in each plot in
Figs.~\ref{fig:11be1-band1-params-scan}--\ref{fig:11be0-band1-params-scan}
represent calculations at successively higher $\Nmax$, obtained for different
oscillator basis length scales given by $\hw$.\footnote{These rotational
  parameters are extracted from the energies of the ``cleanest'' band members,
  least subject to mixing with nearby states.  Thus, the parameters for the
  $K^P=1/2^-$ band in Fig.~\ref{fig:11be1-band1-params-scan} are extracted from
  the three lowest-energy band members ($1/2^-$, $3/2^-$, and $5/2^-$), and
  similarly for the $K^P=1/2^+$ band in Fig.~\ref{fig:11be0-band1-params-scan}.
  On the other hand, for the $K^P=3/2^-$ band, the lower-energy band members are
  in a region of higher level density and subject to mixing with background
  states, which can perturb their energies and make it more difficult to trace
  their evolution across calculations with different $\Nmax$ and $\hw$.
  Therefore, we take energy parameters defined by a straight line through the
  $9/2^-$ and $11/2^-$ band members for the analysis in
  Fig.~\ref{fig:11be1-band2-params-scan} (the rotational fit lines in
  Figs.~\ref{fig:11be1-levels} and~\ref{fig:11be1-energies-nmax} were instead
  obtained as a combined fit to the $3/2^-$, $5/2^-$, $9/2^-$ and $11/2^-$ band
  members).}  Each figure then includes results based on the JISP16~(left) and
Daejeon16~(right) interactions.\footnote{The JISP16
  interaction~\cite{shirokov2007:nn-jisp16} is a two-body interaction derived
  from nucleon-nucleon scattering data by $J$-matrix inverse scattering, then
  adjusted via a phase-shift equivalent transformations to better describe light
  nuclei with $A\leq16$.  The Daejeon16
  interaction~\cite{shirokov2016:nn-daejeon16} is instead obtained from the
  Entem-Machleidt (EM) \nthreelo{} chiral
  interaction~\cite{entem2003:chiral-nn-potl}, softened via a similarity
  renormalization group (SRG) transformation to enhance convergence, and then
  likewise adjusted via a phase-shift equivalent transformation to better
  describe light nuclei with $A\leq16$.}  The $\hw$ range is centered on the
approximate location of the variational energy minimum for the computed ground
state energy, which occurs at $\hw\approx20\,\MeV$ for JISP16 and
$\hw\approx15\,\MeV$ for Daejeon16 (vertical dotted lines).  Experimental values
for the rotational band parameters~\cite{maris2015:berotor2}, extracted from the
observed level energies, are shown for comparison (horizontal
lines).\footnote{\label{fn:11be-expt} The experimental band parameter values for
  the bands in $\isotope[11]{Be}$ are based on fits of the rotational energy formula to the experimental levels, as
  summarized in Table~III of
  Ref.~\cite{maris2015:berotor2}:
    % g1-band1
    for the $1/2^-$ band,
    the $1/2^-$ at $0.320\,\MeV$,
    $3/2^-$ at $2.654\,\MeV$,
    and $5/2^-$ at $3.889\,\MeV$;
    % g1-band2
    for the $3/2^-$ band,
    the $3/2^-$ at $3.955\,\MeV$
    and $5/2^-$ at $5.255\,\MeV$;
    % g0-band1
    for the $1/2^+$ band,
    the $1/2^+$ ground state,
    $3/2^+$ at $3.400\,\MeV$,
    and $5/2^+$ at $1.783\,\MeV$.
    % commentary
    These assignments of levels to bands in $\isotope[11]{Be}$
    follow Refs.~\cite{vonoertzen1997:be-alpha-rotational,bohlen2008:be-band},
    while energies are from Ref.~\cite{npa2012:011}.  However, there are
    conflicting spin-parity assignments in the literature.  For instance, the
    level at $3.4\,\MeV$ was assigned as $3/2^-$ in
    $(t,p)$~\cite{liu1990:11be-tp}, $(3/2^-)$ in $\beta$
    decay~\cite{hirayama2005:11be-beta-delayed}, and $(3/2,5/2)^+$ in
    breakup~\cite{fukuda2004:11be-breakup}, and is evaluated as
    $(3/2^-,3/2^+)$~\cite{npa2012:011}.  The level at $3.9\,\MeV$, was assigned
    as $3/2^+$ in $(t,p)$~\cite{liu1990:11be-tp} but as $5/2^-$ in $\beta$
    decay~\cite{hirayama2005:11be-beta-delayed}, corroborated as negative parity
    in transfer reactions~\cite{bohlen2003:be-transfer-ispun02}, and evaluated
    as $5/2-$~\cite{npa2012:011}.}

The slope parameter $A$ follows entirely from relative energies within a band,
which were already seen from Figs.~\ref{fig:11be1-levels}
and~\ref{fig:11be1-energies-nmax} to be comparatively well-converged.  From the
top panels in
Figs.~\ref{fig:11be1-band1-params-scan}--\ref{fig:11be0-band1-params-scan}, the
calculated $A$ parameter is essentially converged for the Daejeon16 calculations
(in the vicinity of the variational minimum $\hw$), while there is still some
residual dependence on $\Nmax$ (at the few-percent level) and $\hw$ for the
JISP16 calculations.  There is remarkable consistency across these two
interactions, as well as with the experimental values.  A shallower slope
corresponds in the rotational picture to a larger moment of inertia.  Note that
the excited $K^P=3/2^-$ band, by this measure, has a moment of inertia roughly
twice that of the $K^P=1/2^-$ band, both in calculations and experiment (this
greater moment of inertia may be understood in terms
$\alpha$ cluster structure and the molecular
orbitals occupied by the neutrons~\cite{kanadaenyo2012:amd-cluster}).

Even though the Coriolis decoupling parameter $a$
[Figs.~\ref{fig:11be1-band1-params-scan}~(bottom)
  and~\ref{fig:11be0-band1-params-scan}~(middle)] is likewise determined only
from relative energies within a band, it is found to be much more sensitive
to the truncation of the calculation.  (This parameter is extracted essentially
as a second difference in level energies, and numerical second derivatives are
known to be sensitive to uncertainties or fluctuations in the inputs.)  For instance, in the JISP16 calculations for the
$K^P=1/2^-$ band [Fig.~\ref{fig:11be1-band1-params-scan}~(bottom,left)],
although the Coriolis decoupling parameter is deceptively independent of $\Nmax$
at $\hw=20\,\MeV$ (vertical dashed line), there is still a strong $\hw$
dependence, which means that it is not yet possible to extract a converged
value.  On the other hand, $a$ seems to be comparatively well
converged in the Daejeon16 calculations for this same band
[Fig.~\ref{fig:11be1-band1-params-scan}~(bottom,right)] and in close agreement
with experiment ($a\approx0.5$).  For the $K^P=1/2^+$ band, although the $a$
parameter obtained for both interactions is developing a plateau (or shoulder) as a
function of $\hw$, indicative of convergence
[Fig.~\ref{fig:11be0-band1-params-scan}~(middle)], there is still $\Nmax$
dependence at about the $10\%$ level.  The calculated values are consistent with
the much larger decoupling parameter ($a\approx1.8$) experimentally found for this band.

Finally, the excitation energy of the $K^P=3/2^-$ band
[Figs.~\ref{fig:11be1-band2-params-scan}~(bottom)] is poorly converged, as
already found in Sec.~\ref{sec:illustration:nmax}.  The excitation energy of the
unnatural parity $K^P=1/2^+$ band
[Figs.~\ref{fig:11be0-band1-params-scan}~(bottom)] is still highly
$\hw$-dependent (though again deceptively $\Nmax$ independent at $\hw=20\,\MeV$)
for the JISP16 interaction, while the excitation energy obtained in the
Daejeon16 calculation is approaching convergence at the $\sim0.1$--$0.2\,\MeV$
level and appears consistent with experiment.
More detailed comparisons must rely upon extrapolation, as
considered in the following discussion of band parameters along the
$\isotope{Be}$ isotopic chain (Sec.~\ref{sec:params}).

%%%%%%%%%%%%%%%%%%%%%%%%%%%%%%%%%%%%%%%%%%%%%%%%%%%%%%%%%%%%%%%%

\section{Rotational energy parameters for the \boldmath$\isotope{Be}$ isotopes}
\label{sec:params}
%%%%%%%%%%%%%%%%%%%%%%%%%%%%%%%%%%%%%%%%%%%%%%%%%%%%%%%%%%%%%%%%
\begin{figure}[t]
\centerline{\includegraphics[width=1.00\textwidth]{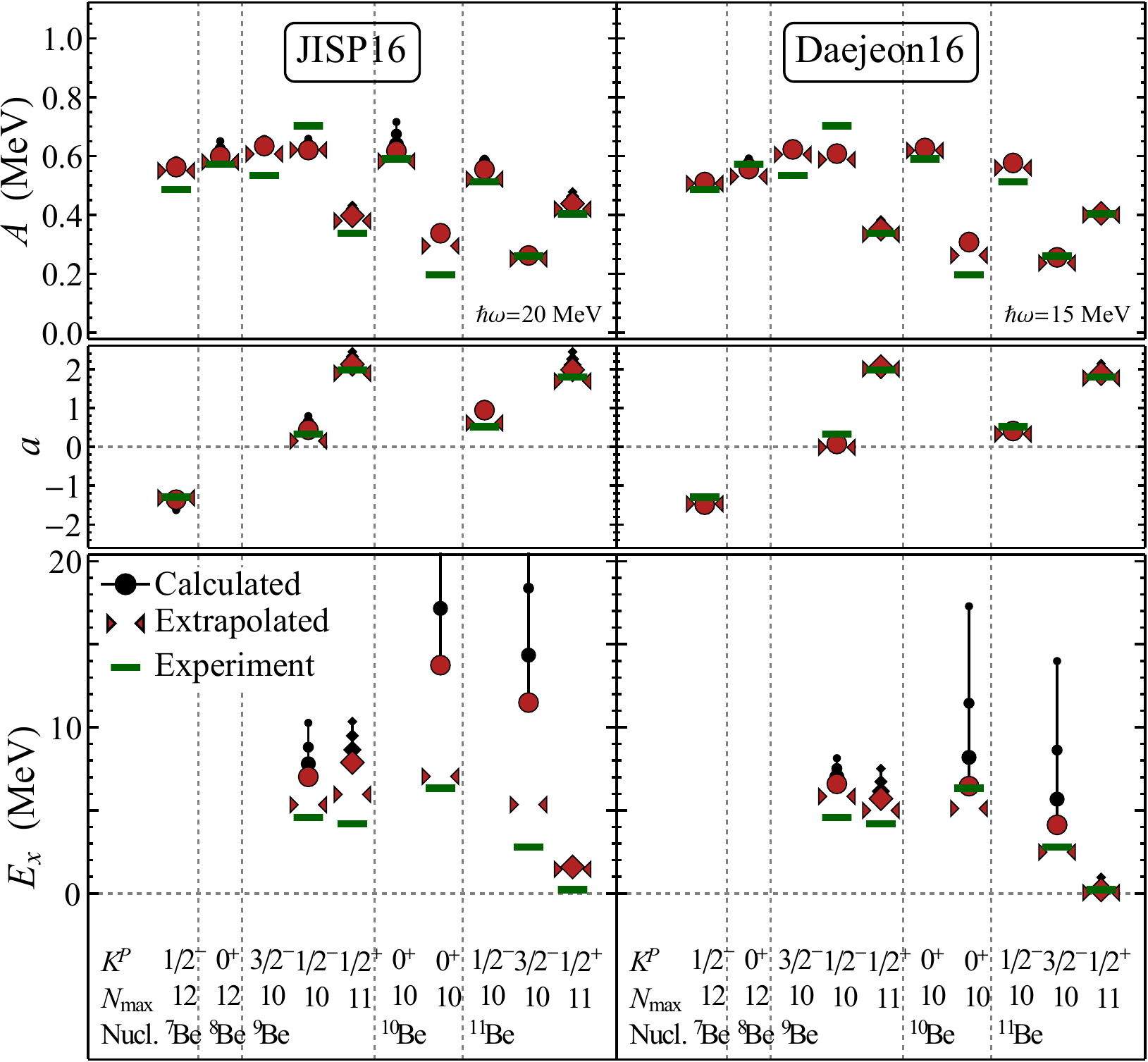}}
\caption{Band energy parameters for $\isotope[7\text{--}11]{Be}$, extracted from
  calculated energy eigenvalues: rotational constant $A$~(top), Coriolis
  decoupling parameter $a$~(middle), and band excitation energy $E_x$~(bottom),
  for the JISP16~(left) and Daejeon16~(right) internucleon
  interactions. Successively larger symbols indicate successively higher $\Nmax$
  values.  Parameter values are also shown based on exponentially extrapolated
  level energies (paired triangles).  Experimental values for the band energy
  parameters (horizontal lines) are shown for
  comparison~\cite{maris2015:berotor2}.  The nuclide, band ($K^P$), and highest
  $\Nmax$ value calculated are noted at the bottom of the plot.  Results are
  obtained from calculations at $\hw=20\,\MeV$ for JISP16 and $\hw=15\,\MeV$ for
  Daejeon16.  }
\label{fig:params}
\end{figure}
%%%%%%%%%%%%%%%%%%%%%%%%%%%%%%%%%%%%%%%%%%%%%%%%%%%%%%%%%%%%%%%%

A variety of rotational bands were identified across the $\isotope{Be}$ isotopes
in Ref.~\cite{maris2015:berotor2-WITH-ERRATUM}.  These include examples of
``short'' bands (terminating at the maximal valence angular momentum) and ``long''
(non-terminating) bands, as well as unnatural parity bands, akin to those
discussed above for $\isotope[11]{Be}$ (Sec.~\ref{sec:illustration}).

We survey the rotational energy parameters extracted from \textit{ab initio}
calculations in Fig.~\ref{fig:params}.  While the calculations in
Ref.~\cite{maris2015:berotor2-WITH-ERRATUM} made use of the JISP16 interaction
without Coulomb contribution, and thus could not be directly compared to
experiment, the present JISP16 and Daejeon16 calculations include Coulomb
interaction and thus may be directly compared to experiment, convergence
permitting.  We do not attempt to display the sensitivity of the extracted
parameters to the basis parameter $\hw$, but rather confine
ourselves to the values obtained at the approximate variational energy minimum
in $\hw$.  However, we do show the sequence of extracted values for four
successive $\Nmax$ truncations, as a more limited indicator of convergence.  We also
show the band parameters obtained from exponentially extrapolated energies.

The first notable feature of the predicted band parameters in
Fig.~\ref{fig:params} is the overall global consistency between predictions with
the JISP16 and Daejeon16 interactions, across the set of bands considered.
Despite the caveat that significant remaining $\hw$-dependence of some of the
extracted band parameters leaves their converged values in doubt (see
Sec.~\ref{sec:illustration:params}), the values for
both the $A$ and $a$ parameters obtained at the variational minimum in $\hw$ are
generally largely $\Nmax$-independent at the $\MeV$ scale considered here.
In contrast, relative excitation energies of different bands are poorly
converged, but even here the extrapolated energies are largely consistent across
interactions.

The overall pattern of rotational band parameters closely
matches experiment.  Where discrepancies arise, the tendency is for the
\textit{ab initio} calculations to be consistent with each other rather than
with experiment.  Here it should be noted that there can be significant
ambiguities in identification of the experimental band members (see,
\textit{e.g.}, footnote~\ref{fn:11be-expt}), as well as fundamental
uncertainties in comparing energies obtained in a bound-state
formalism, such as NCCI, with those from experimental resonant scattering
analysis.

%%%%%%%%%%%%%%%%%%%%%%%%%%%%%%%%%%%%%%%%%%%%%%%%%%%%%%%%%%%%%%%%

\section{Conclusion}

We have explored the dependences of rotational band energy parameters on the
truncation parameters of an oscillator-basis NCCI calculation for the
illustrative case of $\isotope[11]{Be}$
(Figs.~\ref{fig:11be1-band1-params-scan}--\ref{fig:11be0-band1-params-scan})
and, more generally, across the $\isotope{Be}$ isotopes (Fig.~\ref{fig:params}).
We find that \textit{ab initio} calculations can provide quantitatively robust
predictions for rotational band energy parameters in light ($p$-shell) nuclei.
Even subject to the present limitations on \textit{ab initio} many-body
calculations, numerically robust predictions can be made for rotational band
parameters in the $\isotope{Be}$ isotopes.  The results obtained with two
interactions of significantly different pedigree (the JISP16 interaction from
$J$-matrix inverse scattering and the Daejeon16 interaction originating from
chiral perturbation theory) yield highly consistent results.  These results are
also, overall, remarkably consistent with the experimentally observed band
parameters.

%%\clearpage
\section*{Acknowledgements}
%%\sloppy % needed for grants

We thank Jie Chen, Jakub Herko, and Anna McCoy for comments on the manuscript.
This material is based upon work supported by the U.S.~Department of Energy,
Office of Science, under Award Numbers DE-FG02-95ER-40934, DESC00018223
(SciDAC/NUCLEI), and DE-FG02-87ER40371, and by the U.S.~National Science
Foundation under Award Number NSF-PHY05-52843.  This research used computational
resources of the University of Notre Dame Center for Research Computing and of
the National Energy Research Scientific Computing Center (NERSC), a
U.S.~Department of Energy, Office of Science, user facility supported under
Contract~DE-AC02-05CH11231.

%%%%%%%%%%%%%%%%%%%%%%%%%%%%%%%%%%%%%%%%%%%%%%%%%%%%%%%%%%%%%%%%
% bibliography
%%%%%%%%%%%%%%%%%%%%%%%%%%%%%%%%%%%%%%%%%%%%%%%%%%%%%%%%%%%%%%%%

%bibliographystyle{apsrevm}
% Bibliography created with apsrevm.bst
\providecommand{\APSLONG}{}
\providecommand{\ELSEVIER}{}

%bibliography{master,mc,theory,expt,data,books,proc,misc}


\begin{thebibliography}{34}
\expandafter\ifx\csname natexlab\endcsname\relax\def\natexlab#1{#1}\fi
\expandafter\ifx\csname bibnamefont\endcsname\relax
  \def\bibnamefont#1{#1}\fi
\expandafter\ifx\csname bibfnamefont\endcsname\relax
  \def\bibfnamefont#1{#1}\fi
\expandafter\ifx\csname citenamefont\endcsname\relax
  \def\citenamefont#1{#1}\fi
\expandafter\ifx\csname url\endcsname\relax
  \def\url#1{\texttt{#1}}\fi
\expandafter\ifx\csname urlprefix\endcsname\relax\def\urlprefix{URL }\fi
\providecommand{\bibinfo}[2]{#2}
\providecommand{\eprint}[2][arXiv]{\url{#1:#2}}
\renewcommand{\eprint}[2][arXiv]{\url{#1:#2}}

\bibitem{rogers1965:nonspherical-nuclei}
\bibinfo{author}{\bibfnamefont{J.~D.} \bibnamefont{Rogers}},
  \bibinfo{journal}{Annu. Rev. Nucl. Sci.} \textbf{\bibinfo{volume}{15}},
  \bibinfo{pages}{241} (\bibinfo{year}{1965}).

\bibitem{vonoertzen1996:be-molecular}
\bibinfo{author}{\bibfnamefont{W.}~\bibnamefont{von Oertzen}},
  \bibinfo{journal}{Z. Phys. A} \textbf{\bibinfo{volume}{354}},
  \bibinfo{pages}{37} (\bibinfo{year}{1996}).

\bibitem{vonoertzen1997:be-alpha-rotational}
\bibinfo{author}{\bibfnamefont{W.}~\bibnamefont{von Oertzen}},
  \bibinfo{journal}{Z. Phys. A} \textbf{\bibinfo{volume}{357}},
  \bibinfo{pages}{355} (\bibinfo{year}{1997}).

\bibitem{freer2007:cluster-structures}
\bibinfo{author}{\bibfnamefont{M.}~\bibnamefont{Freer}}, \bibinfo{journal}{Rep.
  Prog. Phys.} \textbf{\bibinfo{volume}{70}}, \bibinfo{pages}{2149}
  (\bibinfo{year}{2007}).

\bibitem{pervin2007:qmc-matrix-elements-a6-7}
\bibinfo{author}{\bibfnamefont{M.}~\bibnamefont{Pervin}},
  \bibinfo{author}{\bibfnamefont{S.~C.} \bibnamefont{Pieper}},
  \bibnamefont{and} \bibinfo{author}{\bibfnamefont{R.~B.}
  \bibnamefont{Wiringa}}, \bibinfo{journal}{Phys. Rev. C}
  \textbf{\bibinfo{volume}{76}}, \bibinfo{pages}{064319}
  (\bibinfo{year}{2007}).

\bibitem{bogner2008:ncsm-converg-2N}
\bibinfo{author}{\bibfnamefont{S.~K.} \bibnamefont{Bogner}},
  \bibinfo{author}{\bibfnamefont{R.~J.} \bibnamefont{Furnstahl}},
  \bibinfo{author}{\bibfnamefont{P.}~\bibnamefont{Maris}},
  \bibinfo{author}{\bibfnamefont{R.~J.} \bibnamefont{Perry}},
  \bibinfo{author}{\bibfnamefont{A.}~\bibnamefont{Schwenk}}, \bibnamefont{and}
  \bibinfo{author}{\bibfnamefont{J.}~\bibnamefont{Vary}},
  \bibinfo{journal}{Nucl. Phys. A} \textbf{\bibinfo{volume}{801}},
  \bibinfo{pages}{21} (\bibinfo{year}{2008}).

\bibitem{cockrell2012:li-ncfc}
\bibinfo{author}{\bibfnamefont{C.}~\bibnamefont{Cockrell}},
  \bibinfo{author}{\bibfnamefont{J.~P.} \bibnamefont{Vary}}, \bibnamefont{and}
  \bibinfo{author}{\bibfnamefont{P.}~\bibnamefont{Maris}},
  \bibinfo{journal}{Phys. Rev. C} \textbf{\bibinfo{volume}{86}},
  \bibinfo{pages}{034325} (\bibinfo{year}{2012}).

\bibitem{maris2013:ncsm-pshell}
\bibinfo{author}{\bibfnamefont{P.}~\bibnamefont{Maris}} \bibnamefont{and}
  \bibinfo{author}{\bibfnamefont{J.~P.} \bibnamefont{Vary}},
  \bibinfo{journal}{Int. J. Mod. Phys. E} \textbf{\bibinfo{volume}{22}},
  \bibinfo{pages}{1330016} (\bibinfo{year}{2013}).

\bibitem{dytrych2007:sp-ncsm-dominance}
\bibinfo{author}{\bibfnamefont{T.}~\bibnamefont{Dytrych}},
  \bibinfo{author}{\bibfnamefont{K.~D.} \bibnamefont{Sviratcheva}},
  \bibinfo{author}{\bibfnamefont{C.}~\bibnamefont{Bahri}},
  \bibinfo{author}{\bibfnamefont{J.~P.} \bibnamefont{Draayer}},
  \bibnamefont{and} \bibinfo{author}{\bibfnamefont{J.~P.} \bibnamefont{Vary}},
  \bibinfo{journal}{Phys. Rev. C} \textbf{\bibinfo{volume}{76}},
  \bibinfo{pages}{014315} (\bibinfo{year}{2007}).

\bibitem{dytrych2013:su3ncsm}
\bibinfo{author}{\bibfnamefont{T.}~\bibnamefont{Dytrych}},
  \bibinfo{author}{\bibfnamefont{K.~D.} \bibnamefont{Launey}},
  \bibinfo{author}{\bibfnamefont{J.~P.} \bibnamefont{Draayer}},
  \bibinfo{author}{\bibfnamefont{P.}~\bibnamefont{Maris}},
  \bibinfo{author}{\bibfnamefont{J.~P.} \bibnamefont{Vary}},
  \bibinfo{author}{\bibfnamefont{E.}~\bibnamefont{Saule}},
  \bibinfo{author}{\bibfnamefont{U.}~\bibnamefont{Catalyurek}},
  \bibinfo{author}{\bibfnamefont{M.}~\bibnamefont{Sosonkina}},
  \bibinfo{author}{\bibfnamefont{D.}~\bibnamefont{Langr}}, \bibnamefont{and}
  \bibinfo{author}{\bibfnamefont{M.~A.} \bibnamefont{Caprio}},
  \bibinfo{journal}{Phys. Rev. Lett.} \textbf{\bibinfo{volume}{111}},
  \bibinfo{pages}{252501} (\bibinfo{year}{2013}).

\bibitem{johnson2015:spin-orbit}
\bibinfo{author}{\bibfnamefont{C.~W.} \bibnamefont{Johnson}},
  \bibinfo{journal}{Phys. Rev. C} \textbf{\bibinfo{volume}{91}},
  \bibinfo{pages}{034313} (\bibinfo{year}{2015}).

\bibitem{mccoy2018:spncci-busteni17}
\bibinfo{author}{\bibfnamefont{A.~E.} \bibnamefont{McCoy}},
  \bibinfo{author}{\bibfnamefont{M.~A.} \bibnamefont{Caprio}},
  \bibnamefont{and} \bibinfo{author}{\bibfnamefont{T.}~\bibnamefont{Dytrych}},
  \bibinfo{journal}{Ann. Acad. Rom. Sci. Ser. Chem. Phys. Sci.}
  \textbf{\bibinfo{volume}{3}}, \bibinfo{pages}{17} (\bibinfo{year}{2018}).

\bibitem{kanadaenyo2012:amd-cluster}
\bibinfo{author}{\bibfnamefont{Y.}~\bibnamefont{Kanada-En'yo}},
  \bibinfo{author}{\bibfnamefont{M.}~\bibnamefont{Kimura}}, \bibnamefont{and}
  \bibinfo{author}{\bibfnamefont{A.}~\bibnamefont{Ono}},
  \bibinfo{journal}{Prog. Exp. Theor. Phys.} \textbf{\bibinfo{volume}{2012}},
  \bibinfo{pages}{01A202} (\bibinfo{year}{2012}).

\bibitem{barrett2013:ncsm}
\bibinfo{author}{\bibfnamefont{B.~R.} \bibnamefont{Barrett}},
  \bibinfo{author}{\bibfnamefont{P.}~\bibnamefont{Navr\'{a}til}},
  \bibnamefont{and} \bibinfo{author}{\bibfnamefont{J.~P.} \bibnamefont{Vary}},
  \bibinfo{journal}{Prog. Part. Nucl. Phys.} \textbf{\bibinfo{volume}{69}},
  \bibinfo{pages}{131} (\bibinfo{year}{2013}).

\bibitem{caprio2013:berotor}
\bibinfo{author}{\bibfnamefont{M.~A.} \bibnamefont{Caprio}},
  \bibinfo{author}{\bibfnamefont{P.}~\bibnamefont{Maris}}, \bibnamefont{and}
  \bibinfo{author}{\bibfnamefont{J.~P.} \bibnamefont{Vary}},
  \bibinfo{journal}{Phys. Lett. B} \textbf{\bibinfo{volume}{719}},
  \bibinfo{pages}{179} (\bibinfo{year}{2013}).

\bibitem{maris2015:berotor2-WITH-ERRATUM}
\bibinfo{author}{\bibfnamefont{P.}~\bibnamefont{Maris}},
  \bibinfo{author}{\bibfnamefont{M.~A.} \bibnamefont{Caprio}},
  \bibnamefont{and} \bibinfo{author}{\bibfnamefont{J.~P.} \bibnamefont{Vary}},
  \bibinfo{journal}{Phys. Rev. C} \textbf{\bibinfo{volume}{91}},
  \bibinfo{pages}{014310} (\bibinfo{year}{2015}); \textbf{99}, 029902(E)
  (2019).

\bibitem{caprio2015:berotor-ijmpe}
\bibinfo{author}{\bibfnamefont{M.~A.} \bibnamefont{Caprio}},
  \bibinfo{author}{\bibfnamefont{P.}~\bibnamefont{Maris}},
  \bibinfo{author}{\bibfnamefont{J.~P.} \bibnamefont{Vary}}, \bibnamefont{and}
  \bibinfo{author}{\bibfnamefont{R.}~\bibnamefont{Smith}},
  \bibinfo{journal}{Int. J. Mod. Phys. E} \textbf{\bibinfo{volume}{24}},
  \bibinfo{pages}{1541002} (\bibinfo{year}{2015}).

\bibitem{shirokov2007:nn-jisp16}
\bibinfo{author}{\bibfnamefont{A.~M.} \bibnamefont{Shirokov}},
  \bibinfo{author}{\bibfnamefont{J.~P.} \bibnamefont{Vary}},
  \bibinfo{author}{\bibfnamefont{A.~I.} \bibnamefont{Mazur}}, \bibnamefont{and}
  \bibinfo{author}{\bibfnamefont{T.~A.} \bibnamefont{Weber}},
  \bibinfo{journal}{Phys. Lett. B} \textbf{\bibinfo{volume}{644}},
  \bibinfo{pages}{33} (\bibinfo{year}{2007}).

\bibitem{shirokov2016:nn-daejeon16}
\bibinfo{author}{\bibfnamefont{A.~M.} \bibnamefont{Shirokov}},
  \bibinfo{author}{\bibfnamefont{I.~J.} \bibnamefont{Shin}},
  \bibinfo{author}{\bibfnamefont{Y.}~\bibnamefont{Kim}},
  \bibinfo{author}{\bibfnamefont{M.}~\bibnamefont{Sosonkina}},
  \bibinfo{author}{\bibfnamefont{P.}~\bibnamefont{Maris}}, \bibnamefont{and}
  \bibinfo{author}{\bibfnamefont{J.~P.} \bibnamefont{Vary}},
  \bibinfo{journal}{Phys. Lett. B} \textbf{\bibinfo{volume}{761}},
  \bibinfo{pages}{87} (\bibinfo{year}{2016}).

\bibitem{maris2010:ncsm-mfdn-iccs10}
\bibinfo{author}{\bibfnamefont{P.}~\bibnamefont{Maris}},
  \bibinfo{author}{\bibfnamefont{M.}~\bibnamefont{Sosonkina}},
  \bibinfo{author}{\bibfnamefont{J.~P.} \bibnamefont{Vary}},
  \bibinfo{author}{\bibfnamefont{E.}~\bibnamefont{Ng}}, \bibnamefont{and}
  \bibinfo{author}{\bibfnamefont{C.}~\bibnamefont{Yang}},
  \bibinfo{journal}{Procedia Comput. Sci.} \textbf{\bibinfo{volume}{1}},
  \bibinfo{pages}{97} (\bibinfo{year}{2010}).

\bibitem{aktulga2013:mfdn-scalability}
\bibinfo{author}{\bibfnamefont{H.~M.} \bibnamefont{Aktulga}},
  \bibinfo{author}{\bibfnamefont{C.}~\bibnamefont{Yang}},
  \bibinfo{author}{\bibfnamefont{E.~G.} \bibnamefont{Ng}},
  \bibinfo{author}{\bibfnamefont{P.}~\bibnamefont{Maris}}, \bibnamefont{and}
  \bibinfo{author}{\bibfnamefont{J.~P.} \bibnamefont{Vary}},
  \bibinfo{journal}{Concurrency Computat.: Pract. Exper.}
  \textbf{\bibinfo{volume}{26}}, \bibinfo{pages}{2631} (\bibinfo{year}{2013}).

\bibitem{shao2018:ncci-preconditioned}
\bibinfo{author}{\bibfnamefont{M.}~\bibnamefont{Shao}},
  \bibinfo{author}{\bibfnamefont{H.~M.} \bibnamefont{Aktulga}},
  \bibinfo{author}{\bibfnamefont{C.}~\bibnamefont{Yang}},
  \bibinfo{author}{\bibfnamefont{E.~G.} \bibnamefont{Ng}},
  \bibinfo{author}{\bibfnamefont{P.}~\bibnamefont{Maris}}, \bibnamefont{and}
  \bibinfo{author}{\bibfnamefont{J.~P.} \bibnamefont{Vary}},
  \bibinfo{journal}{Comput. Phys. Commun.} \textbf{\bibinfo{volume}{222}},
  \bibinfo{pages}{1} (\bibinfo{year}{2018}).

\bibitem{ring1980-nuclear-many-body}
\bibinfo{author}{\bibfnamefont{P.}~\bibnamefont{Ring}} \bibnamefont{and}
  \bibinfo{author}{\bibfnamefont{P.}~\bibnamefont{Schuck}},
  \emph{\bibinfo{title}{The Nuclear Many-Body Problem}}
  (\bibinfo{publisher}{Springer-Verlag}, \bibinfo{address}{New York},
  \bibinfo{year}{1980}).

\bibitem{rowe2010:collective-motion}
\bibinfo{author}{\bibfnamefont{D.~J.} \bibnamefont{Rowe}},
  \emph{\bibinfo{title}{Nuclear Collective Motion: Models and Theory}}
  (\bibinfo{publisher}{World Scientific}, \bibinfo{address}{Singapore},
  \bibinfo{year}{2010}).

\bibitem{npa2012:011}
\bibinfo{author}{\bibfnamefont{J.}~\bibnamefont{Kelley}},
  \bibinfo{author}{\bibfnamefont{E.}~\bibnamefont{Kwan}},
  \bibinfo{author}{\bibfnamefont{J.~E.} \bibnamefont{Purcell}},
  \bibinfo{author}{\bibfnamefont{C.~G.} \bibnamefont{Sheu}}, \bibnamefont{and}
  \bibinfo{author}{\bibfnamefont{H.~R.} \bibnamefont{Weller}},
  \bibinfo{journal}{Nucl. Phys. A} \textbf{\bibinfo{volume}{880}},
  \bibinfo{pages}{88} (\bibinfo{year}{2012}).

\bibitem{forssen2008:ncsm-sequences}
\bibinfo{author}{\bibfnamefont{C.}~\bibnamefont{Forssen}},
  \bibinfo{author}{\bibfnamefont{J.~P.} \bibnamefont{Vary}},
  \bibinfo{author}{\bibfnamefont{E.}~\bibnamefont{Caurier}}, \bibnamefont{and}
  \bibinfo{author}{\bibfnamefont{P.}~\bibnamefont{Navratil}},
  \bibinfo{journal}{Phys. Rev. C} \textbf{\bibinfo{volume}{77}},
  \bibinfo{pages}{024301} (\bibinfo{year}{2008}).

\bibitem{maris2009:ncfc}
\bibinfo{author}{\bibfnamefont{P.}~\bibnamefont{Maris}},
  \bibinfo{author}{\bibfnamefont{J.~P.} \bibnamefont{Vary}}, \bibnamefont{and}
  \bibinfo{author}{\bibfnamefont{A.~M.} \bibnamefont{Shirokov}},
  \bibinfo{journal}{Phys. Rev. C} \textbf{\bibinfo{volume}{79}},
  \bibinfo{pages}{014308} (\bibinfo{year}{2009}).

\bibitem{entem2003:chiral-nn-potl}
\bibinfo{author}{\bibfnamefont{D.~R.} \bibnamefont{Entem}} \bibnamefont{and}
  \bibinfo{author}{\bibfnamefont{R.}~\bibnamefont{Machleidt}},
  \bibinfo{journal}{Phys. Rev. C} \textbf{\bibinfo{volume}{68}},
  \bibinfo{pages}{041001} (\bibinfo{year}{2003}).

\bibitem{maris2015:berotor2}
\bibinfo{author}{\bibfnamefont{P.}~\bibnamefont{Maris}},
  \bibinfo{author}{\bibfnamefont{M.~A.} \bibnamefont{Caprio}},
  \bibnamefont{and} \bibinfo{author}{\bibfnamefont{J.~P.} \bibnamefont{Vary}},
  \bibinfo{journal}{Phys. Rev. C} \textbf{\bibinfo{volume}{91}},
  \bibinfo{pages}{014310} (\bibinfo{year}{2015}).

\bibitem{bohlen2008:be-band}
\bibinfo{author}{\bibfnamefont{H.~G.} \bibnamefont{Bohlen}},
  \bibinfo{author}{\bibfnamefont{W.}~\bibnamefont{von Oertzen}},
  \bibinfo{author}{\bibfnamefont{R.}~\bibnamefont{Kalpakchieva}},
  \bibinfo{author}{\bibfnamefont{T.~N.} \bibnamefont{Massey}},
  \bibinfo{author}{\bibfnamefont{T.}~\bibnamefont{Dorsch}},
  \bibinfo{author}{\bibfnamefont{M.}~\bibnamefont{Milin}},
  \bibinfo{author}{\bibfnamefont{{\mbox{Ch}}.}~\bibnamefont{Schulz}},
  \bibinfo{author}{\bibfnamefont{{\mbox{Tz}}.}~\bibnamefont{Kokalova}},
  \bibnamefont{and} \bibinfo{author}{\bibfnamefont{C.}~\bibnamefont{Wheldon}},
  \bibinfo{journal}{J. Phys. Conf. Ser.} \textbf{\bibinfo{volume}{111}},
  \bibinfo{pages}{012021} (\bibinfo{year}{2008}).

\bibitem{liu1990:11be-tp}
\bibinfo{author}{\bibfnamefont{G.-B.} \bibnamefont{Liu}} \bibnamefont{and}
  \bibinfo{author}{\bibfnamefont{H.~T.} \bibnamefont{Fortune}},
  \bibinfo{journal}{Phys. Rev. C} \textbf{\bibinfo{volume}{42}},
  \bibinfo{pages}{167} (\bibinfo{year}{1990}).

\bibitem{hirayama2005:11be-beta-delayed}
\bibinfo{author}{\bibfnamefont{Y.}~\bibnamefont{Hirayama}},
  \bibinfo{author}{\bibfnamefont{T.}~\bibnamefont{Shimoda}},
  \bibinfo{author}{\bibfnamefont{H.}~\bibnamefont{Izumi}},
  \bibinfo{author}{\bibfnamefont{A.}~\bibnamefont{Hatakeyama}},
  \bibinfo{author}{\bibfnamefont{K.~P.} \bibnamefont{Jackson}},
  \bibinfo{author}{\bibfnamefont{C.~D.~P.} \bibnamefont{Levy}},
  \bibinfo{author}{\bibfnamefont{H.}~\bibnamefont{Miyatake}},
  \bibinfo{author}{\bibfnamefont{M.}~\bibnamefont{Yagi}}, \bibnamefont{and}
  \bibinfo{author}{\bibfnamefont{H.}~\bibnamefont{Yanoa}},
  \bibinfo{journal}{Phys. Lett. B} \textbf{\bibinfo{volume}{611}},
  \bibinfo{pages}{239} (\bibinfo{year}{2005}).

\bibitem{fukuda2004:11be-breakup}
\bibinfo{author}{\bibfnamefont{N.}~\bibnamefont{Fukuda}},
  \bibinfo{author}{\bibfnamefont{T.}~\bibnamefont{Nakamura}},
  \bibinfo{author}{\bibfnamefont{N.}~\bibnamefont{Aoi}},
  \bibinfo{author}{\bibfnamefont{N.}~\bibnamefont{Imai}},
  \bibinfo{author}{\bibfnamefont{M.}~\bibnamefont{Ishihara}},
  \bibinfo{author}{\bibfnamefont{T.}~\bibnamefont{Kobayashi}},
  \bibinfo{author}{\bibfnamefont{H.}~\bibnamefont{Iwasaki}},
  \bibinfo{author}{\bibfnamefont{T.}~\bibnamefont{Kubo}},
  \bibinfo{author}{\bibfnamefont{A.}~\bibnamefont{Mengoni}},
  \bibinfo{author}{\bibfnamefont{M.}~\bibnamefont{Notani}},
  \bibinfo{author}{\bibfnamefont{H.}~\bibnamefont{Otsu}},
  \bibinfo{author}{\bibfnamefont{H.}~\bibnamefont{Sakurai}},
  \bibinfo{author}{\bibfnamefont{S.}~\bibnamefont{Shimoura}},
  \bibinfo{author}{\bibfnamefont{T.}~\bibnamefont{Teranishi}},
  \bibinfo{author}{\bibfnamefont{Y.~X.} \bibnamefont{Watanabe}},
  \bibnamefont{and} \bibinfo{author}{\bibfnamefont{K.}~\bibnamefont{Yoneda}},
  \bibinfo{journal}{Phys. Rev. C} \textbf{\bibinfo{volume}{70}},
  \bibinfo{pages}{054606} (\bibinfo{year}{2004}).

\bibitem{bohlen2003:be-transfer-ispun02}
\bibinfo{author}{\bibfnamefont{H.~G.} \bibnamefont{Bohlen}},
  \bibinfo{author}{\bibfnamefont{R.}~\bibnamefont{Kalpakchieva}},
  \bibinfo{author}{\bibfnamefont{W.}~\bibnamefont{von Oertzen}},
  \bibinfo{author}{\bibfnamefont{T.~N.} \bibnamefont{Massey}},
  \bibinfo{author}{\bibfnamefont{B.~G. S.~M.} \bibnamefont{Grimes}},
  \bibinfo{author}{\bibfnamefont{T.}~\bibnamefont{Kokalova}},
  \bibinfo{author}{\bibfnamefont{H.}~\bibnamefont{Lenske}},
  \bibinfo{author}{\bibfnamefont{A.}~\bibnamefont{Lenz}},
  \bibinfo{author}{\bibfnamefont{M.}~\bibnamefont{Milin}},
  \bibinfo{author}{\bibfnamefont{{\mbox{Ch}}.}~\bibnamefont{Schulz}},
  \bibinfo{author}{\bibfnamefont{S.}~\bibnamefont{Thummerer}},
  \bibinfo{author}{\bibfnamefont{S.}~\bibnamefont{Torilov}}, \bibnamefont{and}
  \bibinfo{author}{\bibfnamefont{A.}~\bibnamefont{Tumino}},
  \bibinfo{journal}{Nucl. Phys. A} \textbf{\bibinfo{volume}{722}},
  \bibinfo{pages}{3c} (\bibinfo{year}{2003}).

\end{thebibliography}
\end{document}